\documentclass[useAMS,usenatbib]{mn2e}
\usepackage{stmaryrd}
\usepackage{amssymb}
\usepackage{amsmath}
\usepackage{bbm}
\usepackage{graphicx}

\def\gsim{\;\lower4pt\hbox{${\buildrel\displaystyle >\over\sim}$}\;}
\def\lsim{\;\lower4pt\hbox{${\buildrel\displaystyle <\over\sim}$}\;}
\def\grls{\;\lower4pt\hbox{${\buildrel\displaystyle >\over <}$}\;}

\begin{document}
\title[Intense Magnetic Fields on Magnetars]
{Magnetized massive stars as magnetar progenitors }
\author[R.-Y. Hu and Y.-Q. Lou]{Ren-Yu Hu$^{1}$\thanks{E-mail:
hu-ry07@mails.tsinghua.edu.cn (RYH) and louyq@tsinghua.edu.cn, \ \
lou@oddjob.uchicago.edu
(Y-QL)} and Yu-Qing Lou$^{1}$$^{,2}$$^{,3}$\footnotemark[1]\\
$^{1}$Physics Department and Tsinghua Centre for Astrophysics (THCA),
Tsinghua University, Beijing 100084, China\\
$^{2}$Department of Astronomy and Astrophysics, The University of
Chicago, 5640 South Ellis Avenue, Chicago, IL 60637, USA\\
$^{3}$National Astronomical Observatories, Chinese Academy of
Science, A20, Datun Road, Beijing, 100012, China}

\date{Accepted 2009 February 16. Received 2009 February 06;
in original form 2008 December 16}

\pagerange{\pageref{firstpage}--\pageref{lastpage}} \pubyear{2008}

\maketitle

\begin{abstract}
The origin of ultra-intense magnetic fields on magnetars is a
mystery in modern astrophysics. We model the core collapse dynamics
of massive progenitor stars with high surface magnetic fields in the
theoretical framework of a self-similar general polytropic
magnetofluid under the self-gravity with a quasi-spherical symmetry.
With the specification of physical parameters such as mass density,
temperature, magnetic field and wind mass loss rate on the
progenitor stellar surface and the consideration of a rebound shock
breaking through the stellar interior and envelope, we find a
remnant compact object (i.e. neutron star) left behind at the centre
with a radius of $\sim 10^6$ cm and a mass range of $\sim 1-3$
M$_{\odot}$. Moreover, we find that surface magnetic fields of such
kind of compact objects can be $\sim 10^{14}-10^{15}$ G, consistent
with those inferred for magnetars which include soft gamma-ray
repeaters (SGRs) and anomalous X-ray pulsars (AXPs). The magnetic
field enhancement factor critically depends on the self-similar
scaling index $n$, which also determines the initial density
distribution of the massive progenitor. We propose that magnetized
massive stars as magnetar progenitors based on the
magnetohydrodynamic evolution of the gravitational core collapse and
rebound shock. Our physical mechanism, which does not necessarily
require \textit{ad hoc} dynamo amplification within a fast spinning
neutron star, favours the `fossil field' scenario of forming
magnetars from the strongly magnetized core collapse inside massive
progenitor stars. With a range of surface magnetic field strengths
over massive progenitor stars, our scenario allows a continuum of
magnetic field strengths from pulsars to magnetars. The intense
Lorentz force inside a magnetar may break the crust of a neutron
star into pieces to various extents. Coupled with the magnetar spin,
the magnetospheric configuration of a magnetar is most likely
variable in the presence of exposed convection, differential
rotation, equatorial bulge, bursts of interior magnetic flux ropes
as well as rearrangement of broken pieces of the crust. Sporadic and
violent releases of accumulated magnetic energies and a broken crust
are the underlying causes for various observed high-energy
activities of magnetars.
\end{abstract}


\section{Introduction}

Magnetars are believed to be neutron stars with surface magnetic
field strengths considerably stronger than the quantum critical
value of $B_{\rm QED}=4.4\times10^{13}$ G. There are two main types
of observational manifestations for magnetars: (i) Soft Gamma-ray
Repeaters (SGRs) and (ii) Anomalous X-ray Pulsars (AXPs).
Up to now,
six SGRs and ten AXPs have been identified observationally (see
Mereghetti 2008 for a latest list and an extensive review as well
as very recent powerful explosions of SGR J1550-5418 with a
shortest spin period of 2.07 s). Most recently, a new Galactic
magnetar is reported with very fast optical flares
\citep{Kouveliotou2008, Stefanescu2008, Castro2008}, alluding a
continuum from ordinary dim isolated neutron stars to magnetars.
The ultra-intense surface magnetic fields on magnetars are unique
in the Universe and they are responsible for various high-energy
activities, for example the giant $\gamma$-ray flare of SGR
1806-20 \citep[e.g.][]{Hurley,Palmer}.
Magnetar-like X-ray emissions are also detected from a
rotation-powered pulsar PSR J1846-0258 with an inferred intense
magnetic field of $\sim 4.9\times 10^{13}$ G at the centre of
supernova remnant Kes75 \citep[e.g.][]{Gavriil2008,
Archibald2008}.

Recent observations have also provided clues connecting magnetars
with very massive progenitor stars, for example an infrared
elliptical ring or shell was discovered surrounding the magnetar SGR
1900+14 \citep[e.g.][]{Wachter2008}. However, the formation of
magnetars, especially the origin
of the ultra-intense magnetic field, remains an important open
issue. There are two major contending physical scenarios, viz. the
dynamo scenario versus the fossil-field scenario.

\citet{DT1992} and \citet{TD1993} explored the turbulent dynamo
amplification, occurring primarily in the convection zone of the
progenitor, as well as in a differentially rotating nascent neutron
star, and concluded that very strong magnetic field, in principle up
to $\sim 3\times10^{17}$ G, may be created.
The dynamo mechanism requires an extremely rapid rotation of a
nascent neutron star with a spin period of a few milliseconds.
However, the current population of magnetars appears to be slow
rotators, having spin periods in the range of $\sim 2-12$ s
\citep[e.g.][]{Mereghetti}. Therefore, neutron star dynamo scenario
for magnetars faces a considerable challenge to account for the fact
of slowly rotating magnetars as observed so far.

The fossil-field scenario for the magnetism of compact objects was
first proposed to explain magnetic white dwarfs
\citep[e.g.][]{BS2004, WF2005, FW2005, b12}. It is conceivable
that the magnetic field of white dwarfs may be of fossil origin
from the main-sequence phase of their progenitors, and the attempt
to link magnetic white dwarfs with their main-sequence progenitors
naturally makes the chemically peculiar Ap and Bp stars as
plausible candidates. Observations of \cite{Auriere2003} have
shown that chemically peculiar Ap and Bp stars are generally
magnetic indeed, with a surface magnetic field of $\sim$ 100 G by
Zeeman splittings. In general, magnetic field strengths fall in
the range of $\sim 3\times 10^2-3\times 10^4$ G (e.g. Braithwaite
\& Spruit 2004 and references therein). Magnetic white dwarfs may
be created as a result of rebound shock explosion \citep{b12} and
may further give rise to novel magnetic modes of global stellar
oscillations (Lou 1995). By the magnetic flux conservation during
the stellar evolution, \citet{FW2005} argued that stellar magnetic
fields ($\sim$ 100 G) in their main-sequence phase can be enhanced
up to the range of $\sim 10^6-10^9$ G on the surface of magnetic
white dwarfs. This fossil-field scenario is supported by the
statistics for the mass and magnetic field distributions of
magnetic white dwarfs.

Based on the same scenario and a similar physical argument,
\cite{FW2006} further suggested that the ultra-intense magnetic
field over the surface of magnetars may also come from fossil
magnetic fields. The progenitors of magnetars are expected to be O
stars and early B stars with high surface magnetic fields of $\sim
1000$ G. \cite{FW2008} presented a population synthesis study of the
observed properties of magnetars and found that magnetars arise from
high mass progenitor stars (20M$_{\odot}\lsim$ M $\lsim $
45M$_{\odot}$).

Up to now magnetic fields are directly measured for two O stars,
namely $\theta^1$Ori C ($\sim$ 1 kG, e.g. Donati et al. 2002)
and HD 191612 ($\sim 1.5$ kG, e.g. Donati et al. 2006),
and a couple of early B stars, for example the B0.5V star HD 37061
($\sim$ 650 G, e.g. Hubrig et al. 2006).
\cite{Petit2008a, Petit2008b} carried out systematic
spectropolarimetric observations to search for magnetic fields on
all massive OB stars in the Orion Nebula Cluster star-forming
region. Strong magnetic fields of the order of kG were inferred on
3 stars out of a sample of 8. The existence of strong magnetic
fields on OB stars even appears somewhat overwhelming in contrast
to very few magnetars that have been discovered so far.

With the assumption that neutron stars form during the collapse of
massive progenitors in the Galactic disc with $8\lsim
M/M_{\odot}\lsim 45$ (stellar masses in the main-sequence phase),
and $\sim 8$ percent of massive stars have surface magnetic fields
higher than $\sim 1000$ G, \citet{FW2006} estimated that these
high-field massive progenitors gave birth to 24 neutron stars with
magnetic field $\gsim 10^{14}$ G, consisting a major part of
magnetars. While the fossil-field scenario appears plausible from
the perspective of statistics, it is highly instructive to have a
more direct magnetohydrodynamic (MHD) model description for the
core collapse of high-field massive progenitor stars and to check
whether compact remnants left behind MHD rebound shocks do possess
ultra-intense magnetic fields.

In this paper, we attempt to model magnetized massive progenitor
stars with a quasi-spherical general polytropic magnetofluid under
the self-gravity (Wang \& Lou 2008; Lou \& Hu 2009). We examine
semi-analytic and numerical solutions to explore the self-similar
MHD evolution emerging from dynamic processes of core collapse and
rebound shock travelling in the stellar envelope with a wind mass
loss.
More specifically, we adopt a general polytropic equation of state
(EoS) $p=\kappa(r,\ t)\rho^{\gamma}$ with $p$, $\rho$, $\gamma$, and
$\kappa$ respectively being the gas pressure, mass density,
polytropic index and a proportional coefficient dependent on radius
$r$ and time $t$. Here, $\kappa$ is closely related to the `specific
entropy' and is not necessarily a global constant. By `specific
entropy' conservation along streamlines, another key parameter $q$
arises in self-similar dynamics \citep[see][]{b4}. For $\kappa$
being a global constant, or equivalently $q=0$, the general
polytropic EoS reduces to a conventional polytropic EoS. By further
setting $\gamma=1$, a conventional polytropic gas reduces to an
isothermal gas (e.g. Lou \& Shen 2004).
We also require $\gamma\geqslant 1$ to ensure a positive specific
enthalpy $p/(\gamma-1)$.

\citet{b47} studied the isothermal MHD by including the magnetic
pressure gradient force in the radial momentum equation.
\citet{b48}, Yu, Lou, Bian \& Wu (2006),
\citet{b3} and \citet{b4} generalized the self-similar hydrodynamic
framework by including a completely random transverse magnetic field
with the approximation of a `quasi-spherical' symmetry (e.g.
Zel'dovich \& Novikov 1971); the radial component of such magnetic
field is much weaker than the transverse components.
We conceive a simple `ball of thread' scenario for random
magnetic fields in a massive progenitor star. In other words, a
magnetic field line follows the `thread' meandering within a thin
spherical `layer' in space in a completely random manner. Strictly
speaking, there is always a weak radial magnetic field component
such that field lines in adjacent `layers' remain physically
connected throughout in space. In particular, we emphasize that the
nature of the random magnetic fields inside a progenitor star may be
either those generated by dynamo mechanism probably linked with
convection and differential rotation \citep[e.g.][]{Spruit2002,
Heger2005}, or those of `fossil fields' entrained from molecular
clouds in dynamic processes of star formation \citep[e.g.][]{b3,b4}.

According to numerical simulations of differentially rotating
magnetized stars by Heger et al. (2005),
the dynamo-powered radial magnetic fields in a progenitor star are
about 3 to 4 orders of magnitude weaker than the transverse magnetic
fields during the pre-supernova evolution. In line with this hint,
we simplify the treatment by only dealing with the dominant
transverse field and focus on their dynamic effects on the bulk
motion of gas in the radial direction. By taking the ensemble
average of magnetic fields in each thin spherical `layer', we smooth
out small-scale magnetic field structures and are left with `layers'
of large-scale transverse magnetic fields. We also presume that
small-scale non-spherical flows as a result of the magnetic tension
force may be neglected as compared to large-scale mean radial bulk
flow motions. Therefore on large scales, a completely random
magnetic field contributes to the dynamics in the form of the
average magnetic pressure gradient force and the average magnetic
tension force in the radial direction.

This theoretical model framework of self-similar MHD has been
applied to gravitational core collapse and rebound shock processes
within progenitor stars for supernovae. \cite{b11,b12} modelled the
hydrodynamic and MHD rebound shocks of supernovae in the
self-similar phase of evolution. \cite{HL08b} presented preliminary
results for a shock breakout to reproduce the early X-ray light
curve of supernova SN 2008D \citep[e.g.][]{Soderberg2008,
Mazzali2008}. In this paper, we demonstrate that such a self-similar
MHD process may give birth to a compact remnant with a nuclear
density and a range of ultra-intense surface magnetic fields.
We will see that massive progenitor stars whose collapsing cores
have magnetic fluxes similar to those of magnetars will eventually
collapse into neutron stars with a magnetar level of magnetic fluxes
because of the magnetic flux conservation. In our model scenario,
the neutron star dynamo processes and the required initial rapid
spins of nascent neutron stars may not be necessary.

\section[]{General polytropic self-similar magnetohydrodynamics}

Under the approximation of quasi-spherical symmetry and based on the
physical idea outlined in the introduction, the ideal MHD equations
involve mass conservation, radial momentum equation, specific
entropy conservation along streamlines and the magnetic induction
equation (see Wang \& Lou 2008). We highlight the essential parts of
this formulation of nonlinear MHD equations below.

\subsection[]{Theoretical MHD Model Formulation}

The ideal magnetic induction equation (without the resistivity)
implying the frozen-in condition for the magnetic flux can be cast
into the following form
\begin{equation}
\bigg(\frac{\partial}{\partial t}+u \frac{\partial}{\partial
r}\bigg)(r^2 <B_t^2>)+2r^2<B_t^2>\frac{\partial u}{\partial r}=0\ ,
\label{equ5}
\end{equation}
where $u$ is the bulk radial flow speed and $<B_t^2>$ is the
ensemble mean square of a random transverse magnetic field. The weak
radial component of the magnetic field is determined by equations
(10) and (11) of Yu \& Lou (2005).
With the self-similar MHD
transformation of Wang \& Lou (2008), the ideal MHD equations
together with the magnetic flux frozen-in condition and the general
polytropic EoS can be readily reduced to a set of nonlinear MHD ODEs
in the highly compact form of
\begin{equation}
{\cal X}(x,\alpha,v)\alpha'={\cal A}(x,\alpha,v)\ ,\ \ {\cal
X}(x,\alpha, v)v'={\cal V}(x,\alpha, v)\ ,\label{equ11}
\end{equation}
where the prime $'$ stands for the first derivative $d/dx$, and the
three functionals ${\cal X}$, ${\cal A}$ and ${\cal V}$ are defined
by
\begin{eqnarray}
\!\!\!\!\!\!\!\!&&\!\!\!\!\!\!\!\!\! {\cal X}(x,\ \alpha,\
v)\equiv C
\bigg[2-n+\frac{(3n-2)}{2}q\bigg]\nonumber\\
&&\quad\times\alpha^{1-n+3nq/2} x^{2q}
(nx-v)^q+h\alpha x^2-(nx-v)^2\ ,\nonumber\\
\nonumber\\
\!\!\!\!\!\!\!\!&&\!\!\!\!\!\!\!\!\! {\cal A}(x,\ \alpha,\ v)\equiv
2\frac{x-v}{x}\alpha \big[Cq\alpha^{1-n+3nq/2}
x^{2q}(nx-v)^{q-1}\nonumber\\
&&\quad
+(nx-v)\big]-\alpha\bigg[(n-1)v+\frac{(nx-v)}{(3n-2)}\alpha
+2h\alpha x
\nonumber\\
&&\quad +Cq\alpha^{1-n+3nq/2}x^{2q-1}
(nx-v)^{q-1}(3nx-2v)\bigg]\ , \nonumber\\
\nonumber\\
\!\!\!\!\!\!\!\!&&\!\!\!\!\!\!\!\!\! {\cal V}(x,\ \alpha,\ v)\equiv
2\frac{(x-v)}{x}\alpha\bigg[C
\bigg(2-n+\frac{3n}{2}q\bigg)\nonumber\\
&&\quad\times\alpha^{-n+3nq/2}x^{2q}
(nx-v)^q+hx^2\bigg]\nonumber\\
&&\quad-(nx-v)\bigg[(n-1)v+\frac{(nx-v)}{(3n-2)}\alpha +2h\alpha x
\nonumber\\
&&+Cq\alpha^{1-n+3nq/2}x^{2q-1}(nx-v)^{q-1}(3nx-2v)\bigg]\ .
\label{equ12}
\end{eqnarray}
In this straightforward yet somewhat tedious derivation, we have
useful relations $m=\alpha x^2(nx-v)$ and $q\equiv
2(n+\gamma-2)/(3n-2)$ and we have performed the following MHD
self-similar transformation for a general polytropic gas, viz.
\begin{eqnarray}
&& r=k^{1/2}t^n x,\qquad\qquad u=k^{1/2} t^{n-1} v\ ,\
\nonumber\\
\nonumber\\
&&\rho=\frac{\alpha}{4\pi G t^2}\ ,\qquad\qquad
p=\frac{k t^{2n-4}}{4\pi G}C\alpha^{\gamma}m^q,\ \nonumber\\
\nonumber\\
&&M=\frac{k^{3/2} t^{3n-2} m}{(3n-2)G},\qquad\ <B_t^2>=\frac{k
t^{2n-4}}{G}h\alpha^2x^2,\label{Trans}
\end{eqnarray}
where $G=6.67\times 10^{-8} \hbox{ g}^{-1}\hbox{ cm}^{3}\hbox{
s}^{-2}$ is the gravitational constant, $M$ is the enclosed mass at
time $t$ within radius $r$, $x$ is the independent self-similar
variable, $v(x)$ is the reduced flow speed, $\alpha(x)$ is the
reduced mass density, $m(x)$ is the reduced enclosed mass, $k$, $n$
and $C$ are three parameters, and the dimensionless coefficient $h$
is referred to as the magnetic parameter such that
$<B_{t}^2>=16h\pi^2G\rho^2r^2$. It follows that $q=2/3$ leads to
$\gamma=4/3$ for a relativistically hot gas. Only for the case of
$q=2/3$, can the parameter $C$ be independently chosen; otherwise
for $q\neq 2/3$, $C$ can be set to 1 without loss of generality (Lou
\& Cao 2008; Lou \& Hu 2009). The magnetosonic critical curve (MCC)
is determined by the simultaneous vanishing of the numerator and
denominator on the right-hand sides (RHS) of ODEs (\ref{equ11}) and
(\ref{equ12}).
Once solutions are obtained for $v(x)$ and $\alpha(x)$, the mean
magnetic field strength $<B_{t}^2>^{1/2}$ can be readily
determined from self-similar MHD transformation (\ref{Trans}).
With proper asymptotic conditions as well as eigen-derivatives
across the MCC, nonlinear MHD ODEs (\ref{equ11}) and (\ref{equ12})
can be numerically integrated by using the standard fourth-order
Runge-Kutta scheme \citep[e.g.][]{b6}.

It is straightforward to treat MHD shock conditions for
self-similar solutions to cross the magnetosonic singular surface
${\cal X}(x,\alpha,v)=0$. By the conservations of mass, radial
momentum, and energy as well as magnetic induction equation in the
comoving framework of reference across an MHD shock front, we
obtain a set of jump conditions for an MHD shock that can be cast
into a self-similar form (see Appendix \ref{Ap1}, Wang \& Lou 2008
and Lou \& Hu 2009). Note that the so-called `sound' parameter $k$
in self-similar MHD transformation (\ref{Trans}) is related to the
polytropic sound speed and changes across a shock front, with the
relations $k_2=\lambda^2k_1,\ h_1=h_2=h,\ x_1=\lambda x_2$ where
subscripts $1$ and $2$ refer to the immediate upstream and
downstream sides of a shock and $\lambda$ is a dimensionless
scaling parameter.
Strictly speaking, magnetic fields have
very weak radial components normal to the shock front. Our
treatment of magnetic field coplanar with the shock front
represents a very good approximation for our purposes.

\subsection[]{Analytic Asymptotic MHD Solutions}

The analytic asymptotic solutions at large $x$ of nonlinear coupled
MHD ODEs (\ref{equ11}) and (\ref{equ12}) are
\begin{eqnarray}
\!\!\!\!\!\!\!\!&&\!\!\!\!\!\!\!\!\alpha=Ax^{-2/n}+\cdots, \nonumber\\
\nonumber\\
\!\!\!\!\!\!\!\!&&\!\!\!\!\!\!\!\!v=Bx^{1-1/n}+
\bigg\{-\bigg[\frac{n}{(3n-2)}
+\frac{2h(n-1)}{n}\bigg]A\nonumber\\
&& \qquad +2(2-n)n^{q-1}A^{1-n+3nq/2}\bigg\}x^{1-2/n}+\cdots\ ,
\label{equ23}
\end{eqnarray}
where $A$ and $B$ are two integration constants, referred to as
the mass and velocity parameters \citep[][]{b4}. To ensure the
validity of asymptotic MHD solution (\ref{equ23}), we require
$2/3<n\leqslant 2$; the inequality $n>2/3$ is directly related to
self-similar MHD transformation (\ref{Trans}) where a positive $M$
is mandatory on the ground of physics. For $2/3<n\leqslant 1$,
both mass and velocity parameters $A>0$ and $B$ are fairly
arbitrary. In case of $1<n\leqslant 2$, we should require $B=0$ to
avoid a divergent $v(x)$ at large $x$ unless a flow system is
truncated. Using this asymptotic solution at large $x$ as initial
conditions, the key scaling index $n$ determines the initial mass
density distribution as $\rho\propto r^{-2/n}$. In our
self-similar scenario, the valid range of exponent $n$ corresponds
to a range of density power laws $\rho\propto r^{-3}$ to
$\rho\propto r^{-1}$.

By setting $v=0$ for all $x$ in nonlinear MHD ODEs (\ref{equ11}) and
(\ref{equ12}), we readily obtain an exact global magnetostatic
solution, namely
\begin{equation}
\alpha=A_0x^{-2/n}\ , \label{equ24}
\end{equation}
where the proportional coefficient $A_0$ is given by
\begin{equation}
A_0=\bigg[\frac{n^2-2(1-n)(3n-2)h}
{2(2-n)(3n-2)}n^{-q}\bigg]^{-1/(n-3nq/2)}\ .\label{equ24def}
\end{equation}
This describes a magnetostatic singular polytropic sphere (SPS) with
a substantial generalization of $q\neq 0$; the case of $q=0$ or
$n+\gamma=2$ is included here and corresponds to a conventional
polytropic gas in magnetostatic equilibrium. A further special case
of $n=\gamma=1$ corresponds to a magnetostatic singular isothermal
sphere (SIS). Physically, expression (\ref{equ24def}) requires
$q\neq2/3$ and $h<h_c\equiv n^2/[2(1-n)(3n-2)]$ for the existence of
the global magnetostatic SPS solution in a general polytropic gas.
For $n=4/5$, $h_c$ reaches the minimum value $h_c=4$. This places a
constraint only when $n<1$; while for $n\geqslant1$, parameter $h>0$
is fairly arbitrary.

There exists an analytic asymptotic MHD solution approaching the
magnetostatic SPS solution at small $x$ (referred to as the
`quasi-magnetostatic' asymptotic solution), namely $v=Lx^K$ and
$\alpha=A_0x^{-2/n}+Nx^{K-1-2/n}$ where $K$ is the root of the
following quadratic equation
\[
\bigg[\frac{n^2}{2(3n-2)}+nh+\frac{(3n-2)}{2}Q\bigg]
\bigg[K^2+\frac{(3n-4)}nK\bigg]
\]
\begin{equation}\label{qstatic1}
\ +\frac{2(2-n)(1-n)}{n}h +\frac{n^2+(3n-2)^2(1-4/n)Q}{(3n-2)}=0\ ,
\end{equation}
where $Q\equiv q\big\{n^2/[2(2-n)(3n-2)]-(1-n)h/(2-n)\big\}$ is
introduced for notational clarity (Lou \& Wang 2006, 2007). In a
certain regime, at least one root of quadratic equation
(\ref{qstatic1}) satisfies $\Re (K)>1$ and therefore
quasi-magnetostatic solutions do exist. The two coefficients $L$ and
$N$ are simply related by the following algebraic expression
\begin{equation}
n(K-1)N=(K+2-2/n)A_0L\ .\label{qstatic2}
\end{equation}
In this case, the magnetic Lorentz force (i.e. magnetic pressure and
tension forces together) and the gas pressure force are in the same
order of magnitude in the regime of small $x$.

It can be proved that the parameter regimes where
quasi-magnetostatic solutions exist are $\gamma\geqslant1$, $h<h_c$,
$q<2/3$, $n<0.8$ and $\gamma\geqslant1$, $h<h_c$, $q>2/3$. With
parameters outside these two regimes, the so-called strong-field
asymptotic solutions at small $x$, for which the magnetic force
dominates over the thermal pressure force, have been shown to exist
\citep{b49,b3,b4}.

\section[]{Formation of Compact Magnetars}

\subsection[]{Model Progenitors and Compact Remnants}

To be specific, the radial range of our model solutions is set
within $r_{\rm i}<r<r_{\rm o}$, where $r_{\rm i}=10^6$ cm if the
compact object is a neutron star or black hole, and $r_{\rm
o}=10^{12}$ cm as the radius of a typical massive main-sequence
star \citep[e.g.][]{Herrero,SH}. Massive stars may have undergone
tremendous mass losses before the onset of gravitational core
collapses, and a progenitor immediately before a core collapse may
have already lost the entire hydrogen envelope and become a
compact Wolf-Rayet star with a radius of $\sim 10^{11}$ cm and a
mass of $\sim 4-8M_{\odot}$ (see e.g. Soderberg et al. 2008 and
Mazzali et al. 2008 for recent observations). We attempt to
further identify plausible conditions of forming magnetars by
gravitational core collapses of massive progenitor stars.

At the initial time of our analysis, the model should approximately
represent the final stage of a massive progenitor star before the
gravitational core collapse. The mass density, temperature and
magnetic field on the progenitor stellar surface can be inferred
from the observed range. From these quantities, one estimates the
dimensionless magnetic parameter $h$, which plays an important role
in the MHD evolution of
quasi-magnetostatic solutions. 
By self-similar MHD transformation (\ref{Trans}) and the ideal gas
law, we have
\begin{equation}
Ck^{1-3q/2}=\frac{k_B T}{\mu (4\pi
G)^{\gamma-1}G^q(3n-2)^q\rho^{\gamma-1}M^q}\ ,\label{equk}
\end{equation}
where $T$ is the gas temperature, $k_B$ is Boltzmann's constant,
and $\mu$ is the mean molecular (atomic) weight for gas particles.
For a gas mainly of ionized hydrogen and a stellar mass of several
solar masses, one can estimate parameter $k$ according to
expression (\ref{equk}) from $\gamma$ and the local density $\rho$
and temperature $T$. For typical values of $\rho\sim 10^{-5}$ g
cm$^{-3}$ and $T\sim10^5-10^6$ K on the stellar surface of a
massive progenitor in the late phase (just before the
gravitational core collapse) and $\gamma=1.3$ and $n=0.7$, we
estimate a range of $k\sim 10^{16}-10^{17}$ cgs unit.

With $t\rightarrow 0^+$ and/or $r\rightarrow\infty$ (i.e. for the
regime of large $x$), self-similar MHD solutions follow asymptotic
form (\ref{equ23}). For such analytic asymptotic solutions, the
mass density simply scales as
\begin{equation}
\rho=\frac{Ak^{1/n}}{4\pi G}r^{-2/n}\ \label{Aden}
\end{equation}
and is independent of time $t$. The reason to refer $A$ as the mass
parameter is now apparent. With asymptotic mass density scaling
(\ref{Aden}), one can estimate $A$ from the surface mass density of
a progenitor star. We have the radial flow velocity
\begin{equation}
u=Bk^{1/(2n)}r^{1-1/n}\ ,\label{Vden}
\end{equation}
also independent of $t$. The radial flow velocity near the stellar
surface relates to the mass accretion rate or mass loss rate by
\begin{equation}
\dot{M}=-4\pi r^2\rho u\ .\label{MLoss}
\end{equation}
With expressions (\ref{Vden}) and (\ref{MLoss}), one can determine
velocity parameter $B$ from the stellar mass loss rate for $B>0$
or mass accretion rate for $B<0$. Observationally, mass loss rates
of Galactic OB stars fall in the range of $\sim 10^{-5}-10^{-7}$
M$_{\odot}$ yr$^{-1}$ \citep[e.g.][]{Lamers1993}. Mass loss rates
of Wolf-Rayet stars are much higher than massive main-sequence
stars, falling in the range of $\sim 10^{-4}-10^{-6}$ M$_{\odot}$
yr$^{-1}$ \citep[e.g.][]{Singh1986}. Very recently,
\citet{Puls2008} provide an extensive review on mass loss rates of
massive stars.

In summary, with the mass density, temperature, magnetic field and
mass loss rate specified at the surface of a massive progenitor
star, one can determine all parameters of asymptotic solution
(\ref{equ23}) and integrate nonlinear MHD ODEs (\ref{equ11}) and
(\ref{equ12}) inwards to produce an MHD profile for the progenitor
interior and envelope. We still have the freedom to choose the
rebound shock position after the gravitational core collapse.
Physically, the speed of the rebound shock depends on the core
collapse, for example the EoS and the neutrino reheating, etc.
Here we simply treat it as an adjustable parameter to search for
downstream solutions within the shock front as $x\rightarrow 0^+$.

With $t\rightarrow\infty$ and/or $r\rightarrow0^+$ (i.e. for the
regime of small $x$), the final evolution may approach either a
quasi-magnetostatic solution (Lou \& Wang 2007) or the
strong-field solutions (Wang \& Lou 2007, 2008). It is desirable
that after a long lapse (i.e. $t\rightarrow\infty$), the enclosed
mass within $r_{\rm i}$ evolves towards a constant value and form
a compact remnant of nuclear mass density. Asymptotically, the
enclosed mass takes the form of
\begin{equation}
M=\frac{nk^{1/n}A_0}{(3n-2)G}r^{3-2/n} \label{MassQS}
\end{equation}
for quasi-magnetostatic solutions and becomes independent of $t$.
This appears consistent with the scenario of forming a central
compact object with a strong magnetic field \citep[][]{b12}. In a
companion paper, we will show that the asymptotic enclosed mass
for strong-field solutions depends on $t$, and may be invoked to
model a continuous accretion or outflow around a nascent neutron
star (Hu \& Lou 2009, in preparation). As quasi-magnetostatic
solutions require that $h<h_c$, we arrive at an interesting
situation, i.e. in order to give birth to a stable neutron star
with an ultra-intense magnetic field, the massive progenitor star
needs to be magnetized but not too much. We shall see presently
that $h<h_c$ is generally satisfied for massive main-sequence
stars.

We now introduce the outer initial mass $M_{\mbox{o,ini}}$ and the
inner ultimate mass $M_{\mbox{i,ult}}$ in the same manner as done in
\cite{b11,b12}
and regard them as rough estimates for the the initial progenitor
mass and the final mass of the remnant compact object,
respectively. The ratio of these two masses is given by
$M_{\mbox{o,ini}}/M_{\mbox{i,ult}}
=\lambda^{*}(r_{\mbox{o}}/r_{\mbox{i}})^{(3-2/n)}$ with
$\lambda^{*}\equiv (A/A_0)\lambda^{-2/n}$.
The ratio of outer initial magnetic field at the surface of the
progenitor star to the inner final magnetic field of the central
compact remnant is
$<B_{\mbox{o,ini}}^2>^{1/2}/<B_{\mbox{i,ult}}^2>^{1/2}=\lambda^{*}
(r_{\mbox{o}}/r_{\mbox{i}})^{(1-2/n)}$. The factor $\lambda^{*}$
is insignificant as compared to the radial variation of magnetic
field, i.e. the very radial dependence of $r^{(1-2/n)}$. As
$n\rightarrow 2/3$ and the polytropic index $\gamma$ approaches
$4/3$, this scaling approaches $r^{-2}$. With
$r_{\mbox{o}}=10^{12}$ cm and $r_{\mbox{i}}=10^6$ cm, the magnetic
field strength can be rapidly enhanced by a factor up to $\sim
10^{12}$. Thus, for a magnetar (i.e. neutron star) to have a
surface magnetic field strength of $<B_{\mbox{i,ult}}^2>^{1/2}\sim
10^{15}$G, we need a magnetic field of $\sim 10^3$G over the
progenitor stellar surface, which is attainable for magnetic
massive OB stars.

\subsection[]{Numerical Model Calculations}

The initial time to apply our solutions is estimated by the time
when the rebound shock crosses $r_{\rm i}$, viz. $t_1=[r_{\rm
i}/(k^{1/2}x_s)]^{1/n}$ \citep[][]{b11}. Here we suppose that
roughly from $t_1$ on the collapse and rebound shock inside the
progenitor star have already evolved into a self-similar phase
(typically this process takes a few milliseconds). The rebound
shock travels outwards through the stellar interior into the
envelope in $\sim 10^4-10^6$ s \citep[e.g.][]{b12,LH08, HL08a}. We
set $t_2$ as the time when the rebound shock reaches $r_{\rm o}$,
roughly around the shock breakout. We will also show self-similar
MHD solutions at $t_{m1}=1$ s as an intermediate time between
$t_1$ and $t_2$, and at $t=\infty$.

\begin{figure*}
 \includegraphics[width=\textwidth]{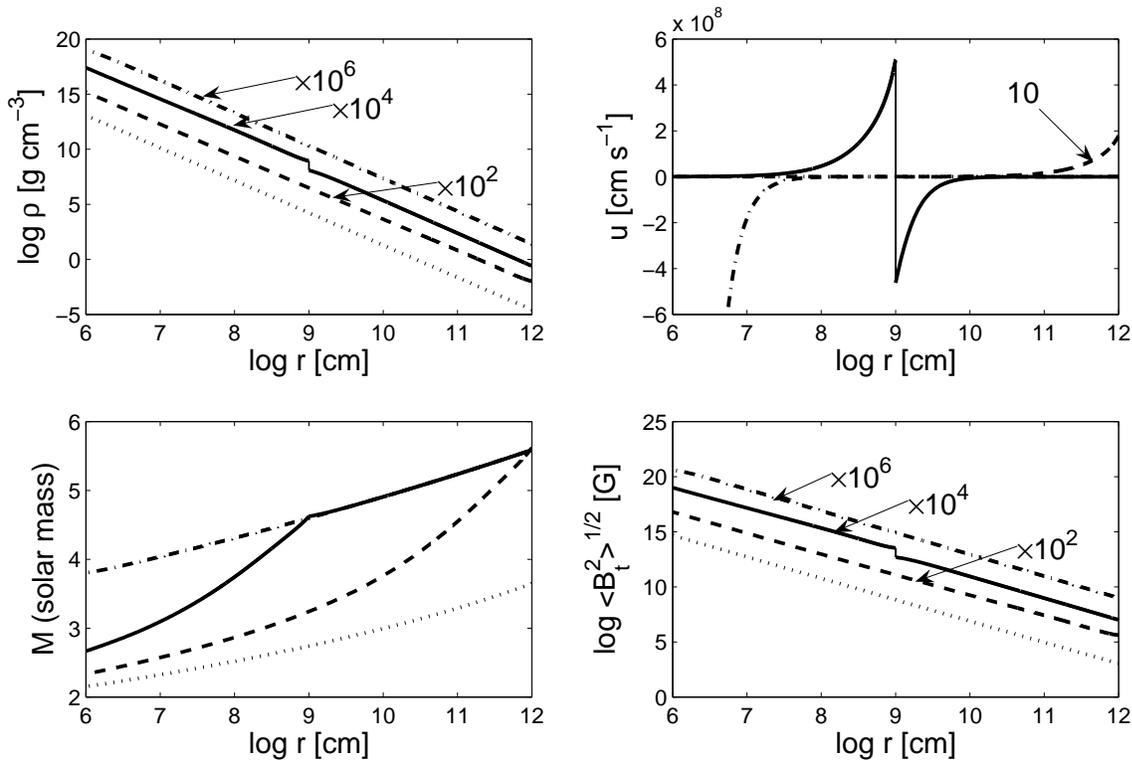}
 \caption{
 The radial profiles of mass density, radial flow velocity, enclosed mass and
 transverse magnetic field at four different epochs. For all panels, the
 dash-dotted, solid, dashed and dotted curves are at $t_1$, $t_{m1}$, $t_2$,
 and $t=\infty$, respectively. Mass density, velocity and magnetic
 field profiles are multiplied by numerical factors marked along the
 curves for a compact presentation.}
 \label{Evol}
\end{figure*}

As an example of illustration, we choose $n=0.673,\ q=0,\
\gamma=1.327$ for a conventional polytropic gas. From the analysis
in subsection 3.1, solutions with a smaller $n$ ($>2/3$) tend to
give a larger $M_{\mbox{i,ult}}$ and $<B_{\mbox{i,ult}}^2>^{1/2}$.
The choice of parameters here is a compromise among multiple
numerical tests. With a typical surface mass density of $2.5\times
10^{-5}$ g cm$^{-3}$, a surface temperature $3\times 10^4$ K, a
mass loss rate $10^{-6}$ M$_{\odot}$ yr$^{-1}$, and a surface
magnetic field strength $10^3$ G for a massive progenitor star, we
estimate $k_1=1.55\times 10^{16}$ cgs unit, $A=8.4378$,
$B=1.27\times10^{-7}$ and $h=1.52\times10^{-4}$. Such parameter
$h$ ensures that a stellar core collapse evolves into a
quasi-magnetostatic manner for small $x$. Here, inequality $h<h_c$
is readily satisfied for this set of adopted parameters. We set a
rebound shock reaching a radius $r=10^9$ cm at $t=1$ s, which
fixes the shock location and travel speed. With these parameters,
a global self-similar rebound shock solution can be constructed
and the temporal evolution is shown in Figure \ref{Evol} at
sampled epochs $t_1=3.49\times10^{-5}$ s, $t_{m1}$,
$t_2=2.87\times10^4$ s and $t=\infty$.


The initial progenitor mass is around 5.59 M$_{\odot}$, consistent
with observations of \cite{Soderberg2008}.
At epoch $t_1$ (dash-dotted curve), the rebound shock has
not yet emerged. The radial velocities around the surface of the
proto-neutron star point inwards, corresponding to a core collapse
process leading to a subsequent rebound shock travelling outwards
inside the progenitor envelope. Meanwhile, the outer part of the
progenitor envelope still flows slowly outwards for a stellar wind.
Such kind of self-similar shock manifestation with a collapsing
inner part and an expanding outer part is made possible from the
quasi-magnetostatic asymptotic solutions \citep[see][]{b11} and is
another form of envelope expansion with core collapse (EECC)
proposed by Lou \& Shen (2004). At epoch $t_{m1}$ (solid curve), the
outgoing rebound shock emerges and travels within the stellar
envelope. Figure \ref{Evol} clearly shows this discontinuity across
the outgoing MHD shock front. The rebound shock evolves in a
self-similar manner with an outgoing speed decreasing with time $t$
for $n<1$ (see Lou \& Hu 2008 and Hu \& Lou 2008b for a comparison
with numerical simulations).
The immediate downstream side of the shock has an outward velocity
and the immediate upstream side has an inward velocity. The
enclosed mass of the downstream side decreases rapidly towards the
centre while that of the upstream side remains nearly unchanged.
Across the shock front, both mass density and magnetic field
strength are enhanced by a factor of 6.96. It can be derived that
$<B_t^2>^{1/2}_1/<B_t^2>^{1/2}_2=\rho_1/\rho_2=2/[(\gamma+1){\cal
M}_1^2]+(\gamma-1)/(\gamma+1)$ where ${\cal M}_1$ is the upstream
Mach number in the comoving shock framework of reference. The
maximum enhancement across the shock is
$(\gamma+1)/(\gamma-1)=7.12$ for our adopted value of polytropic
index $\gamma=1.327$.

This rebound shock breaks out from the stellar envelope in
$t_2\sim 3\times 10^4$ s. We see at that moment the flow velocity
within the spherical volume previously occupied by the progenitor
star becomes very much reduced in the wake of the rebound shock,
and the gas there gradually approaches the quasi-magnetostatic
phase of evolution.
Coupled with radiation mechanisms, for instance the thermal
bremsstrahlung of hot electrons with a temperature $T\gsim 10^7$
K, and using the dynamic profiles shown in Figure \ref{Evol}, one
may compute the radiation detected and reproduce the X-ray or
$\gamma$-ray light curves observed
\citep[e.g.][]{HL08b,Soderberg2008,Mazzali2008}. Eventually, flow
velocities of the entire system tend to zero and the enclosed
masses at all radii remain unchanged. From Figure \ref{Evol}, we
see that for the initial and final stages of the MHD evolution,
the mass density and magnetic field distributions obey power laws,
consistent with the asymptotic analysis in subsection 3.1.
Finally, within radius $r_{\rm i}$ of the inner compact remnant,
the enclosed mass is 2.15 M$_{\odot}$ with a mean density of
$1.02\times 10^{15}$ g cm$^{-3}$ for a neutron star. As the mean
surface magnetic field strength is $<B_{\mbox{i,ult}}^2>^{1/2}\sim
4.70\times10^{14}$ G, this neutron star should be indeed regarded
as a magnetar.

\begin{figure}
 \includegraphics[width=0.5\textwidth]{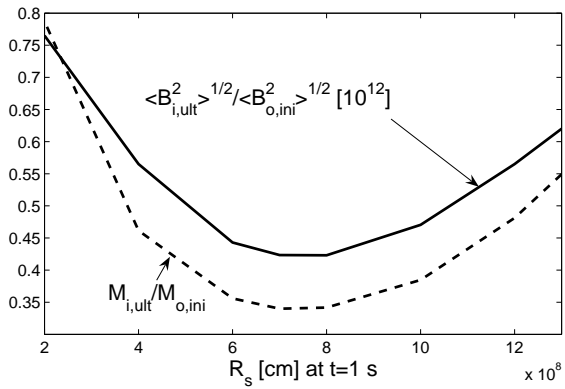}
 \caption{
 The dependence of the ensemble averaged magnetic field strength
 enhancement factor $<B_{\mbox{i,ult}}^2>^{1/2}/<B_{\mbox{o,ini}}^2>^{1/2}$
 and the mass ratio $M_{\mbox{i,ult}}/M_{\mbox{o,ini}}$ on the MHD rebound shock
 radii for $n=0.673,\ q=0,\ \gamma=1.327$, a surface mass density
 $2.5\times10^{-5}$ g cm$^{-3}$, a surface temperature $3\times10^4$ K,
 and a mass loss rate $10^{-6}$ M$_{\odot}$ yr$^{-1}$.} \label{Dep_Rs}
\end{figure}

\begin{figure}
 \includegraphics[width=0.5\textwidth]{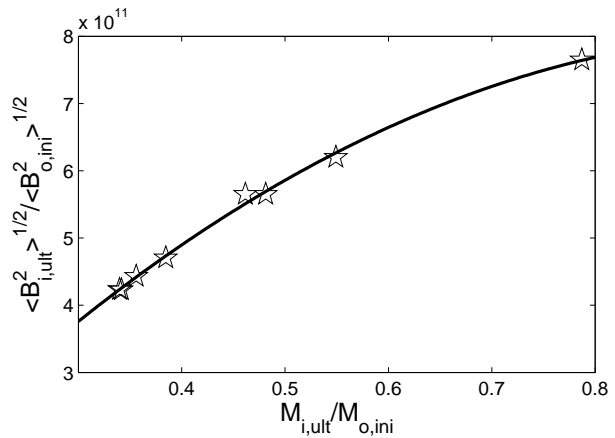}
 \caption{
The ensemble averaged magnetic field strength enhancement
$<B_{\mbox{i,ult}}^2>^{1/2}/<B_{\mbox{o,ini}}^2>^{1/2}$ versus the
mass ratio $M_{\mbox{i,ult}}/M_{\mbox{o,ini}}$ after a long time
of rebound shock evolution. The star points mark our numerical
explorations with different shock properties in the case of
 $n=0.673,\ q=0,\ \gamma=1.327$, a surface mass density $2.5\times10^{-5}$ g
cm$^{-3}$, a surface temperature $3\times10^4$ K and a mass loss
rate $10^{-6}$ M$_{\odot}$ yr$^{-1}$. Our numerical data can be
best fitted (solid curve) with a quadratic relation
$<B_{\mbox{i,ult}}^2>^{1/2}/<B_{\mbox{o,ini}}^2>^{1/2}=-8.8\times10^{11}
(M_{\mbox{i,ult}}/M_{\mbox{o,ini}})^2+1.8\times10^{12}
(M_{\mbox{i,ult}}/M_{\mbox{o,ini}})-7.1\times10^{10}$.}
 \label{Dep_BM}
\end{figure}

Numerical explorations indicate that the ultimate magnetic field
on the neutron star is proportional to the initial magnetic field
on the progenitor stellar surface. However, the magnetic
enhancement factor
$<B_{\mbox{i,ult}}^2>^{1/2}/<B_{\mbox{o,ini}}^2>^{1/2}$ and the
mass ratio $M_{\mbox{i,ult}}/M_{\mbox{o,ini}}$ indeed depend on
model parameters and in particular, on shock properties and
self-similar scaling indices $n$ and $q$. As long as the density
scales as $r^{-2/n}$, index $n$ must be set to approach the
limiting value 2/3 to ensure a sufficiently massive progenitor
star. In Figure \ref{Dep_Rs}, we plot these two ratios versus the
selected rebound shock radii at $t=1$ s. The two curves suggest
that a shock with a medium travel speed is associated with a
minimum magnetic field enhancement. The magnetic enhancement
factor appears grossly proportional to the mass ratio. We
illustrate the relation between these two ratios in Figure
\ref{Dep_BM}. Note that the magnetic enhancement factor does not
exceed $10^{12}$.

\begin{figure}
 \includegraphics[width=0.5\textwidth]{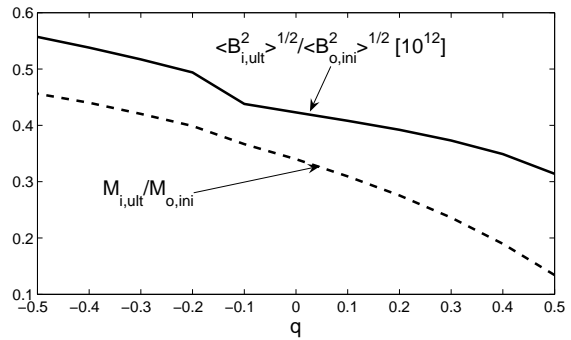}
 \caption{ The dependence of the ensemble averaged magnetic field strength
 enhancement factor $<B_{\mbox{i,ult}}^2>^{1/2}/<B_{\mbox{o,ini}}^2>^{1/2}$
 and the mass ratio $M_{\mbox{i,ult}}/M_{\mbox{o,ini}}$ on scaling index
 parameter $q$ in the case of $n=0.673$, a surface mass density
 $2.5\times10^{-5}$ g cm$^{-3}$, a surface temperature $3\times 10^4$ K, a
 mass loss rate $10^{-6}$ M$_{\odot}$ yr$^{-1}$ and a shock position
 $r_s=7\times10^8$ cm at $t=1$ s. }
 \label{Dep_Q}
\end{figure}

We further examine the influence of parameter $q$ for a general
polytropic magnetofluid shown in Fig. \ref{Dep_Q}. With a larger
$q$, both the mass ratio and magnetic enhancement factor become
less. The qualitative behaviours of quasi-magnetostatic solutions
in general polytropic cases are similar to the conventional
polytropic cases.

Our self-similar MHD rebound shock model analysis suggests that
should there be a continuum of stellar surface magnetic field
strengths over massive progenitor stars, there is then a
corresponding continuum from normal radio pulsars to magnetars
in terms of magnetic field strengths over the compact stellar
surface. The key factors that decide whether a remnant neutron
star possesses an ultra-intense magnetic field include scaling
indices $n$ and $q$, the initial surface magnetic field of
progenitor star, and the strength (or speed) of the rebound shock.
Our analysis also resolves the difficulty posed by the survey of
\citet{Petit2008b}. That is, a massive progenitor star with a
proper range of surface magnetic field strengths is most likely
but not necessarily leads to a magnetar after the gravitational
core collapse and the emergence of a rebound shock. There are
additional constraints to be satisfied. Both Figures \ref{Dep_Rs}
and \ref{Dep_Q} demonstrate that the magnetic enhancement factor
cannot exceed $\sim 10^{12}$, and may be as low as $\sim 10^{11}$
even as $n\rightarrow 2/3$. This implies that the conditions for
producing magnetars are fairly strict. Another major limit arises
as the fact that the magnetic enhancement factor correlates to the
mass ratio. If this mass ratio is too high, the compact remnant
mass may exceed the Tolman-Oppenheimer-Volkoff (TOV) limit ($\sim
3-3.2$ M$_{\odot}$; Rhoades \& Ruffini 1974),
and the core object would collapse further to form a black hole.
The upper bound for the mass of neutron stars then places a limit
on their surface magnetic field strengths.

\section[]{Conclusions and Discussion}

We combine semi-analytic and numerical self-similar MHD solutions to
model the gravitational core collapse, the rebound shock explosion
of a magnetized massive progenitor star, and the formation of a
central remnant compact magnetar. As natural extension and
generalization to recent supernova rebound shock models of Lou \&
Wang (2006, 2007), we invoked quasi-magnetostatic asymptotic
solutions and asymptotic solutions far away from the centre for a
{\it general polytropic magnetofluid}. With the magnetic frozen-in
condition imposed, the surface magnetic field of a nascent neutron
star can be very much stronger than that
of its progenitor by a factor of $\sim 10^{11}-10^{12}$ during
processes of the gravitational core collapse and the self-similar
rebound shock breakout.
Therefore if the progenitor is a magnetic massive star with a
surface magnetic field strength of $\sim 10^3$ G, it would have a
good chance to produce a magnetar at the centre of its supernova
remnant.
Here we propose that magnetars may be produced through powerful
supernova explosions of magnetized massive progenitor stars. Such
physical origin is also supported by statistical inferences from
observational surveys \citep[e.g.][]{FW2006}.
While the magnetic flux of the collapsing core inside the
progenitor star may come from either main-sequence stellar dynamo
processes or `fossil fields' of molecular clouds, it will be
dragged and squeezed into the newborn neutron star by the
conservation of magnetic fluxes. At least for magnetic massive
stars as magnetar progenitors, the post-supernova dynamo processes
inside the remnant neutron star may not be necessary for producing
the ultra-intense surface magnetic field. Magnetic field strengths
in the interior of such magnetars thus produced are expected to be
even stronger and are gravitationally buried and confined by the
nuclear-density matter.

If the surface magnetic field strength of a massive progenitor
happens to be even higher, possibly up to $\sim 3\times 10^4$ G or
more,
a nascent compact magnetar may then possess an ultra-intense surface
magnetic field up to $\sim 10^{15}-3\times 10^{16}$ G or higher.
Such strong magnetic fields can give rise to various stellar
activities and magnetic reconnection can release stored magnetic
energy sporadically and violently (e.g. Low \& Lou 1990). For
example, if one approximates the magnetospheres of SGRs, AXPs, and
radio pulsars as in gross force-free equilibria (i.e. electric
currents parallel to magnetic fields), then the magnetic energies
retained in such magnetospheric systems are higher than the
corresponding potential field configurations with the same footpoint
magnetic field at the stellar surface. A loss of equilibrium most
likely triggered by magnetic reconnections may then lead to
outbursts of available magnetic energies. In the solar and stellar
contexts, such processes correspond to solar/stellar flare
activities and coronal mass ejections.
For compact stellar object like neutron stars, this type of
dramatic magnetic energy releases might fuel `magnetic fireballs'
which produce short-hard gamma-ray bursts, such as those reported
giant flares \citep[e.g.][]{Hurley} as well as very recent
outbursts of SGR J1550-5418 and SGR 1627-41.
After giant flares, a
neutron star may still recover surface magnetic fields of order of
$\sim 10^{14}-10^{15}$ G manifested as either an AXP or a SGR.

For an intense magnetic field buried inside a magnetar, there
is yet another possible source of activities. The magnetic Lorentz
force may break the crust of a neutron star into pieces to various
levels. Coupled with the magnetar spin, a broken crust can give rise
to various activities. For examples, chunks of crust may pile up
around the equatorial bulge and rearrange themselves to generate
stellar seismic activities in a random manner; interior magnetic
flux ropes may burst into the magnetar magnetosphere randomly at
weak points of a crust; if the crust is more or less destroyed by
the Lorentz force, then footpoints of magnetic field lines can be
moved around by convective motions and possible differential
rotations of a magnetar; the manoeuvre of magnetic footpoints over
the surface of a magnetar leads to variable configurations of the
magnetosphere and thus produces magnetic activities including
analogues of `flares' and `coronal mass ejections' mentioned
earlier.

Regarding observed magnetar-like X-ray emissions from a
rotation-powered pulsar PSR J1846-0258
inside supernova remnant Kes75 \citep[e.g.][]{Gavriil2008,
Archibald2008}, our model can accommodate this pulsar resulting from
a magnetized massive progenitor yet with a lower surface magnetic
field strength. Such magnetar-like activities are physically
associated more with strong magnetospheric field strengths. With the
current observational evidence, it appears not necessary to
postulate that magnetars evolve from fast spinning radio pulsars.

In our `ball of thread' scenario for a random magnetic field
within a massive progenitor star, the large-scale mean of such a
random magnetic field is idealized as dominantly transverse with a
fairly weak radial component. By the approximation of
quasi-spherical symmetry, low-amplitude small-scale deviations,
oscillations or fluctuations are randomly distributed about the
mean flow profile and are expected to co-evolve with large-scale
MHD profiles (e.g. Lou \& Bai 2009 in preparation; Cao \& Lou 2009
in preparation). During the processes of the gravitational core
collapse and the MHD rebound shock breakout, a remnant neutron
star forms with high nuclear density and ultra-intense magnetic
field, while a major portion of the interior and envelope of the
massive progenitor star is driven out into the interstellar space
by the rebound shock with entrained strong magnetic field. Our
semi-analytic model describes a large-scale self-similar MHD
evolution for the supernova explosion of a magnetized massive
progenitor. Along with this large-scale evolution, the central
magnetar and the thrown-out stellar materials may certainly follow
their own courses of adjustment or rearrangement (e.g. Lou 1994).
For example, considerable radial components of magnetic field
should emerge in a random turbulent manner. In fact, magnetars are
expected to possess magnetospheres with various possible
configurations (see Low \& Lou 1990 for constructing force-free
stellar magnetic field configurations). By numerical simulations
with `fossil fields', \cite{BS2004} illustrated examples of
processes for magnetic field rearrangement to occur within a few
Alfv\'en timescales. The emergent stable structure of magnetic
field for magnetic Ap stars appears always of `offset dipole' type
(with complex and twisted magnetic field configurations inside),
consistent with observations. By this analogy, we presume that
such magnetic field rearrangement processes could take place very
rapidly during the formation of intensely magnetized neutron
stars, whose Alfv\'en timescale is of the order of $\sim 0.1$ s.
Eventually, a magnetar can possess a variety of magnetic field
configurations (e.g. Low \& Lou 1990).

Magnetars observed so far appear to be slow rotators, while
massive stars are in general rapid rotators with typical equatorial
speeds of $\sim 200$ km s$^{-1}$ (e.g. Fukuda 1982). Therefore,
significant angular momentum transfer may have taken place in the
stellar evolution of magnetic massive stars. \cite{Spruit1999,
Spruit2002} has shown that magnetic fields can be created in stably
stratified layers inside a differentially rotating star.
Heger et al. (2005) gave detailed rotating stellar evolution
calculations for stars in the mass range of $\sim 12-35$ M$_{\odot}$
incorporating the dynamo-powered magnetic field.
In general, it is found that magnetic breaking decreases the final
spin rate of the collapsing iron core by a factor of $\sim 30-50$
when compared with the nonmagnetic case. The `fossil' (or
primordial) magnetic fields may have similar dynamic effects
regarding the re-distribution of angular momentum inside a massive
star. In particular, for magnetar formation, high magnetic fields
may lead to stronger core-envelope coupling during the hydrogen and
helium burning phase of the SN progenitor, and the collapsing iron
core and the compact supernova remnant is expected to be even slower
rotators. This is in accordance with the population of slowly
rotating magnetars observed.

In our model at this stage of development, the stellar rotation
is not included to simplify the mathematical treatment.
Conceptually, it could be possible to design an axisymmetric MHD
problem to accommodate stellar differential rotation in order to
explore the re-distribution of angular momentum during the processes
of gravitational core collapse, MHD rebound shock as well as
collimated MHD outflows from polar regions with shocks (e.g. shocked
MHD jets). The overall magnetic field configuration could be
predominantly toroidal but a relatively weak radial magnetic field
component is necessary to exert an effective magnetic torque to
break or slow down the stellar core rotation. As the core materials
rapidly collapse towards the centre under the gravity, the
mechanical angular momentum is transferred outwards in an outgoing
envelope with shock. Along the rotation axis, collimated outflows or
jets may emerge to breakthrough the polar stellar envelope and part
of the mechanical angular momentum is carried outwards by rotating
polar collimated outflows. For a semi-analytic self-similar approach
to this time-dependent problem, one might be able to perform a
self-similar transformation combining time $t$ with two spatial
coordinates, say $r$ and $\theta$. It might be possible to derive
asymptotic solutions in the regime of slow rotators for this
two-dimensional magnetar formation problem. One also expects the
existence of several MHD singular surfaces in deriving flow
solutions. Physically, such a scheme if tractable semi-analytically
and/or numerically can be applied to a wide range of gravitational
collapses of rotating systems, including magnetars, pulsars,
magnetic white dwarfs, protostars, planets and so forth.

In addition to the quasi-magnetostatic asymptotic solutions adopted
and exemplified in this paper, it may be possible that a magnetized
massive progenitor star evolves towards the strong-field asymptotic
solutions ultimately \citep{b49,b3,b4}, involving a material
fall-back process towards the central remnant neutron star. Under
this situation, the enclosed mass within a certain radius keeps
increasing until the mass of the neutron star exceeds the TOV limit.
Such an MHD fall-back process offers a possible means to form
stellar mass black holes as compact remnants in supernova and
hypernova explosions. We emphasize that such a mechanism requires a
strong magnetic field inside a progenitor and the magnetic force
becomes dominant during the fall-back process.

\section*{Acknowledgments}

This research was supported in part by Tsinghua Centre for
Astrophysics (THCA), by the National Natural Science Foundation of
China (NSFC) grants 10373009 and 10533020 and by the National
Basic Science Talent Training Foundation
(NSFC J0630317) at Tsinghua University, and by the SRFDP
20050003088 and 200800030071 and the Yangtze Endowment from the
Ministry of Education at Tsinghua University. The hospitality of
Institut f\"ur Theoretische Physik und Astrophysik der
Christian-Albrechts-Universit\"at Kiel Germany and of
International Center for Relativistic Astrophysics Network
(ICRANet) Pescara, Italy is gratefully acknowledged.

\appendix

\section[]{MHD Shock Conditions}

In the shock comoving framework of reference, the MHD shock jump
conditions in terms of the reduced self-similar variables are
\begin{eqnarray}
&&\alpha_1(nx_1-v_1)=\lambda\alpha_2(nx_2-v_2)\ ,\label{equ181}\\
\nonumber\\
&&\qquad C\alpha_1^{2-n+3nq/2}x_1^{2q}(nx_1-v_1)^q\nonumber\\
&&\qquad\quad +\alpha_1(nx_1-v_1)^2
+\frac{h\alpha_1^2x_1^2}{2}\nonumber\\
&&\qquad\qquad
=\lambda^2\bigg[C\alpha_2^{2-n+3nq/2}x_2^{2q}(nx_2-v_2)^q
\nonumber\\
&&\qquad\qquad\quad
+\alpha_2(nx_2-v_2)^2+\frac{h\alpha_2^2x_2^2}{2}\bigg]\ ,
\label{equ182}\\
\nonumber\\
&&
\frac{2\gamma}{(\gamma-1)}C\alpha_1^{1-n+3nq/2}x_1^{2q}(nx_1-v_1)^q
\nonumber\\
&&\qquad\qquad\qquad +(nx_1-v_1)^2+2h\alpha_1x_1^2
\nonumber\\
&& \qquad\quad =\lambda^2\bigg[\frac{2\gamma}{(\gamma-1)}C
\alpha_2^{1-n+3nq/2}x_2^{2q}(nx_2-v_2)^q\nonumber\\
&&\qquad\qquad\qquad\qquad +(nx_2-v_2)^2+2h\alpha_2x_2^2\bigg]\
\label{equ183}
\end{eqnarray}
\citep{b4}. Once we have $(x_2,\ \alpha_2,\ v_2)$ on the immediate
downstream side of a shock front, we can obtain $(x_1,\ \alpha_1,\
v_1)$ explicitly for the immediate upstream side using MHD shock
relations (\ref{equ181})$-$(\ref{equ183}) \citep{b4, LH08} or vice
versa. In the case of $q=2/3$, there are only two independent
relations among equations (\ref{equ181})$-$(\ref{equ183}) and we
could choose parameter $\lambda>0$ arbitrarily. Hence we can set
$k_1=k_2$ or $\lambda=1$ in this situation. This treatment will
not alter the relations of the resulting dimensional physical
variables. In general, the outgoing travel speed of a rebound
shock varies with time $t$ for $n\neq 1$: shock accelerates for
$n>1$, shock speed remains constant for $n=1$, and shock
decelerates for $n<1$. \label{Ap1}
\end{document}